\documentclass[runningheads,fleqn]{svmult}
\usepackage{makeidx}   
\usepackage{graphicx}  
\usepackage{subeqnar}  
\usepackage{multicol}  
\usepackage{taphys}    
\makeindex             
\begin{document}
\title*{Theory of Collective Dynamics in Multi-Agent Complex Systems}
\toctitle{Focusing of a Parallel Beam to Form a Point
\protect\newline in the Particle Deflection Plane}
%
%
\titlerunning{Theory of Collective Dynamics in Multi-Agent Complex Systems}
%
\author{Neil F. Johnson\inst{1,3}
\and Sehyo C. Choe\inst{1}
\and Sean Gourley\inst{1}
\and Timothy Jarrett\inst{1}
\and Pak Ming Hui\inst{2}}
\authorrunning{Neil F. Johnson et al.}
%
%
\institute{Physics Department, Clarendon Laboratory, Oxford University, U.K.
\and  Department of Physics, Chinese University of Hong Kong, Hong Kong
\and Corresponding author: n.johnson@physics.ox.ac.uk}

\maketitle              

\begin{abstract} We discuss a crowd-based theory for describing the collective
behavior in Complex Systems comprising multi-agent populations competing
for a limited resource. These systems -- whose binary versions we refer to as
B-A-R (Binary Agent Resource) systems -- have a dynamical evolution which is
determined by the aggregate action of the heterogeneous, adaptive agent
population. Accounting for the strong correlations between agents' strategies,
yields an accurate analytic description of the system's dynamics.
\end{abstract}

\section{Introduction} Complex Systems -- together with their dynamical behavior
known as Complexity -- are thought to pervade much of the natural, informational,
sociological, and economic world
\cite{casti1,book}.  Complex Systems are probably
best defined in terms of a list of common features which distinguish them
from `simple' systems, and from systems which are just `complicated' as opposed to
being complex. A list of Complex System
`stylized facts' should include: feedback and adaptation at the macroscopic
and/or microscopic level, many (but not too many) interacting parts,
non-stationarity, evolution, coupling with the environment, and observed dynamics
which depend upon the particular realization of the system.

Casti has argued that  \cite{casti1} `.... a decent mathematical formalism to
describe and analyze the [so-called] El Farol Problem would go a long way toward
the creation of a viable theory of complex, adaptive systems'. The rationale
behind this statement is that the El Farol Problem, which was originally proposed
by Brian Arthur \cite{farol} to demonstrate the essence of Complexity in financial
markets involving many interacting agents, incorporates the key features of a
Complex System in an everyday setting. Very briefly, the El Farol Problem concerns
the collective decision-making of a group of potential bar-goers (i.e. agents) who
repeatedly try to predict whether they should attend a potentially overcrowded bar
on a given night each week. They have no information about the others' predictions.
Indeed the only information available to each agent is global, comprising a string
of outcomes (`overcrowded' or `undercrowded') for a limited number of previous
occasions. Hence they end up having to predict the predictions of others. No
`typical' agent exists, since all such typical agents would then make the same
decision, hence rendering their common prediction scheme useless. With the
exception of Ref.
\cite{uselfarol}, the physics literature has focused on a simplified
binary form of the El Farol Problem called the Minority Game (MG) as introduced by
Challet and Zhang
\cite{origMG,webpage}. 

In this paper, we present a theoretical framework for describing a class of
Complex Systems comprising competitive multi-agent populations, which we
refer to as B-A-R (Binary Agent Resource) systems. The resulting Crowd-Anticrowd
theory is not limited to MG-like games, even though we focus on MG-like games in
order to demonstrate the accuracy of the  approach. The theory is built around
the correlations or `crowding' in strategy-space. Since the theory only makes
fairly modest assumptions about a specific game's dynamical behavior, it can
describe the dynamics in a wide variety of systems comprising competitive
populations
\cite{lanl}.

\begin{figure}
\includegraphics[width=.7\textwidth]{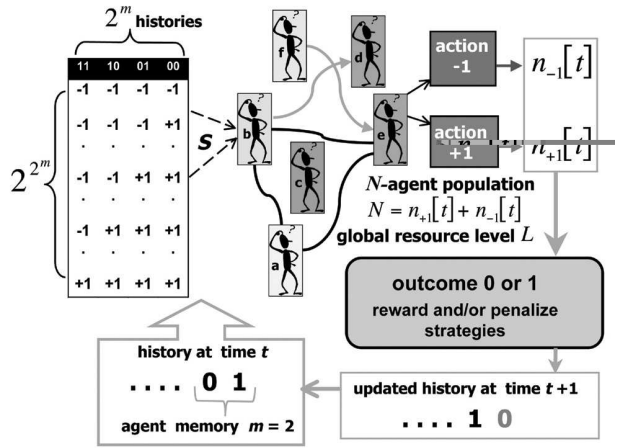}
\caption{Schematic representation of B-A-R (Binary Agent Resource) system. At
timestep
$t$, each agent decides between action $-1$ and action $+1$ based on the
predictions of the $S$ strategies that he possesses. A total of $n_{-1}[t]$ agents
choose
 $-1$, and $n_{+1}[t]$ choose $+1$. Agents may be
subject to some underlying network structure which may be static or evolving, and
ordered or disordered (see Refs. \cite{lanl,sean}). The agents' actions are
aggregated, and a global outcome 0 or 1 is assigned. Strategies are
rewarded/penalized one virtual point according to whether their predicted action
would have been a winning/losing action.}
\label{figure1}
\end{figure}

\section{B-A-R (Binary Agent Resource) Systems}
Figure 1 summarizes the generic form of
the B-A-R (Binary Agent Resource) system under consideration. At timestep
$t$, each agent (e.g. a bar customer, a commuter, or a market agent) decides
whether to enter a game where the choices are action $+1$ (e.g. attend the bar,
take route A, or buy) and action $-1$ (e.g. go home, take route B, or sell). The
global information available to the agents is a common memory of the most recent
$m$ outcomes, which are represented as either 0 (e.g. bar attendance below seating
capacity
$L$) or 1 (e.g. bar attendance above seating capacity $L$). Hence this outcome
history is represented by a binary bit-string of length
$m$. For general
$m$, there will be
$P=2^{m}$ possible history bit-strings. These history bit-strings can
alternatively be represented in decimal form: $\mu =\{0,1,...,P-1\}$. For
$m=2$, for example, 
$\mu =0$\ corresponds to $00$, $\mu =1$\ corresponds to 01 etc.  A strategy
consists of a predicted action, $-1$ or $+1$, for each possible history
bit-string. Hence there are
$2^{P=2^m}$ possible strategies.

\begin{figure}
\includegraphics[width=.7\textwidth]{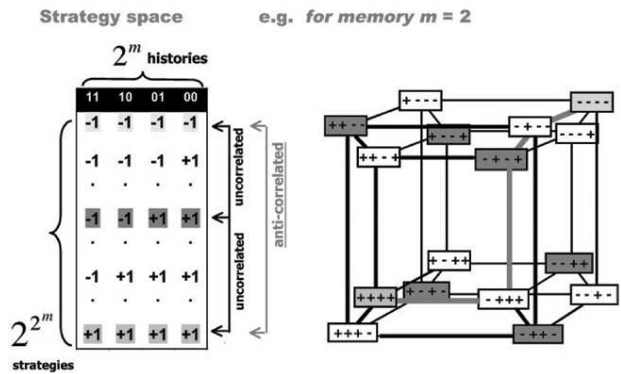}
\caption{Strategy Space for $m=2$, together with some example strategies (left).
The strategy space shown is known as the Full Strategy Space (FSS) and contains all
possible permutations of the actions $-1$ and $+1$ for each history. There are
$2^{2^{m}}$ strategies in the FSS. The $2^{m}$ dimensional hypercube (right) shows
all 
$2^{2^{m}}$ strategies in the FSS at its vertices. The shaded strategies form a
Reduced Strategy Space RSS. There are $2.{2^{m}}=2P$ strategies in the RSS.
The grey shaded line connects two strategies with a Hamming distance separation of
4. }
\label{figure2}
\end{figure}

Figure 2 shows the $m=2$ strategy space from Figure 1.  A strategy is a set of
instructions to describe what action an agent should take, given any particular
history
$\mu$. The strategy
space is the set of strategies from which agents are allocated their strategies.
The strategy space shown is known as the Full Strategy Space
(FSS) and contains all possible permutations of the actions
$-1$ and $+1$ for each history. As such there are $2^{2^{m}}$ strategies in this
space. One can choose a subset of $2.2^m$ strategies, called a Reduced
Strategy Space (RSS), such that any pair within this subset has one of the
following two characteristics: (i) {\em Anti-correlated}. For example, any two
agents using the ($m=2$) strategies
$(-1,-1,+1,+1)$ and $(+1,+1,-1,-1)$ respectively, would take the opposite action
irrespective of the sequence of previous outcomes and hence the history. Their net
effect on the excess demand
$D[t]=n_{+1}[t]-n_{-1}[t]$ (which is an important quantity in a socio-economic
setting such as a financial market) therefore cancels out at each timestep,
regardless of the history. Hence they will not contribute to fluctuations in
$D[t]$. (ii) {\em Uncorrelated}. For example, any two agents
using the  strategies 
$(-1,-1,-1,-1)$ and $(-1,-1,+1,+1)$ respectively, would take the opposite action
for two of the four histories, and the same action for the remaining two
histories. If the histories occur equally often, the actions of the two
agents will be uncorrelated on average. Note that the strategies
in the RSS can be labeled by $R=\{1,2,...,2P=2.2^m\}$. 

The strategy allocation among agents can be described in terms of a tensor
$\Omega $ describing the distribution of strategies among
the $N$ individual agents. This strategy allocation is typically fixed from
the beginning of the game, hence acting as a quenched disorder in the system. The
rank of the tensor $\Omega $\ is given by the number of strategies $S$ that each
agent holds. We note that a single
$\Omega$ `macrostate' corresponds to many possible `microstates' describing the
specific partitions of strategies among the agents. Hence the present
Crowd-Anticrowd theory retained at  the level of a given $\Omega$, describes the
set of all games which belong to that same $\Omega$ macrostate.  We also note that
although
$\Omega$ is not symmetric, it can be made so since the agents do not
distinguish between the order in which the two strategies are picked. Given this,
we will henceforth focus on $S=2$ and consider the symmetrized version of the
strategy allocation matrix given by
${\Psi}=\frac{1}{2}({\Omega}+{\Omega}^T)$. 

\section{Crowd-Anticrowd Formalism} 
Consider an arbitrary timestep $t$ during a run
of the game. We will focus on evaluating the `excess demand' $D[t]\equiv
D\left[
\underline{S}[t],\mu
\lbrack t]\right]=n_{+1}[t]-n_{-1}[t]$, although any other function of
$n_{+1}[t]$ and $n_{-1}[t]$ can be evaluated in a similar way. Here
$\underline{S}[t]$ is the $2P$-dimensional score-vector whose $R$'th component is
the virtual point score for strategy $R$. [Strategies gain/lose one virtual point
at each timestep, according to whether their predicted action would have been a
winning/losing action]. The current history is $\mu\lbrack t]$. The
standard deviation of
$D[t]$ for this given run, corresponds to a time-average for a given realization
of 
$\Psi
$ and a given set of initial conditions. Summing over the RSS, we have:
$D\left[ \underline{S}[t],\mu \lbrack t]\right]
=\sum_{R=1}^{2P}a_{R}^{\mu
\lbrack t]}n_{R}^{\underline{S}[t]}$. 
The quantity $a_{R}^{\mu \lbrack t]}=\pm 1$ is the
action predicted by strategy $R$ in response to the history bit-string $\mu $\ at
time
$t$.  The quantity
$n_{R}^{\underline{S}[t]}$ is the number of agents using strategy $R$ at time $t$.
We use the notation $\left\langle X[t]\right\rangle _{t}$ to denote a time-average
over the variable $X[t]$ for a given $\Psi $. Hence
\begin{equation}
\left\langle D\left[ \underline{S}[t],\mu \lbrack t]\right] \right\rangle _{t}
=\sum_{R=1}^{2P}\left\langle a_{R}^{\mu \lbrack
t]}n_{R}^{\underline{S}[t]}\right\rangle _{t} =\sum_{R=1}^{2P}\left\langle
a_{R}^{\mu \lbrack t]}\right\rangle _{t}\left\langle
n_{R}^{\underline{S}[t]}\right\rangle _{t}  \nonumber
\end{equation}
where we have used the property that $a_{R}^{\mu \lbrack t]}$ and
$n_{R}^{\underline{S}[t]}$ are uncorrelated. We now consider the special case in
which all histories are visited equally on average: even if this situation
does not hold for a specific $\Psi$, it may indeed hold once the averaging over 
$\Psi$ has also been taken. For example, in the Minority Game all histories are
visited equally at small
$m$ and a given $\Psi$. If  we take
the additional average over all $\Psi$, then the same is also true for large
$m$.  Under the property of equal histories: 
\begin{eqnarray}
\left\langle D\left[ \underline{S}[t],\mu \lbrack t]\right] \right\rangle _{t}
&=&\sum_{R=1}^{2P}\left( \frac{1}{P}\sum_{\mu =0}^{P-1}a_{R}^{\mu \lbrack
t]}\right) \left\langle n_{R}^{\underline{S}[t]}\right\rangle _{t}  \label{hist} \\
&=&\sum_{R=1}^{P}\left( \frac{1}{P}\sum_{\mu =0}^{P-1}a_{R}^{\mu \lbrack
t]}+a_{\overline R}^{\mu \lbrack
t]}\right) \left\langle n_{R}^{\underline{S}[t]}\right\rangle
_{t} = \sum_{R=1}^{P}0.\left\langle n_{R}^{\underline{S}[t]}\right\rangle _{t}
\nonumber
\\ &=&0  \nonumber 
\end{eqnarray}
where we have used the exact result that $a_{R}^{\mu \lbrack
t]}=-a_{\overline R}^{\mu \lbrack t]}$ for all
$\mu[t]$, and the approximation  
$\left\langle n_{R}^{\underline{S}[t]}\right\rangle _{t}=\left\langle n_{\overline
R}^{\underline{S}[t]}\right\rangle _{t}$. This approximation is reasonable for a
competitive game since there is typically no a priori best strategy: if the
strategies are distributed fairly evenly among the agents, this then implies that
the average number playing each strategy is approximately equal and hence
$\left\langle n_{R}^{\underline{S}[t]}\right\rangle _{t}=\left\langle n_{\overline
R}^{\underline{S}[t]}\right\rangle _{t}$.  In the event that all  histories are
not equally visited over time, even after averaging over all $\Psi$, it may still
happen that the system's dynamics is restricted to equal visits to some {\em
subset} of histories. 
In this case one can then carry out the averaging in Equation (\ref{hist}) over
this subspace of histories. More generally, the averagings in this formalism can be
carried out with appropriate frequency weightings for each history. In fact, any
non-ergodic dynamics can be incorporated if one knows the appropriate history path
\cite{sean}.

The variance of the excess demand $D[t]$ is given by
\begin{equation}
\sigma _{\Psi}^{2} =\left\langle D\left[ \underline{S}[t],\mu \lbrack t]
\right] ^{2}\right\rangle _{t}-\left\langle D\left[ \underline{S}[t],\mu
\lbrack t]\right] \right\rangle _{t}^{2}  \ .
\end{equation} For simplicity, we will here assume the game output is unbiased and
hence $\left\langle D\left[
\underline{S}[t],\mu
\lbrack t]\right] \right\rangle _{t}=0$. Hence
\begin{equation}
\sigma _{\Psi}^{2} =\left\langle D\left[
\underline{S}[t],\mu \lbrack t]\right] ^{2}\right\rangle _{t}  
=\sum_{R,R^{\prime }=1}^{2P}\left\langle a_{R}^{\mu \lbrack
t]}n_{R}^{\underline{S}[t]}a_{R^{\prime }}^{\mu \lbrack t]}n_{R^{\prime
}}^{\underline{S}[t]}\right\rangle _{t}.  \nonumber
\end{equation} 
Using the identities \underline{$a_{R}$}$.
\underline{a_{R^{\prime }}}=P$\ (fully correlated), \underline{$a_{R}$}$.
\underline{a_{R^{\prime }}}=-P$\ (fully anti-correlated), and \underline{$
a_{R}$}$.\underline{a_{R^{\prime }}}=0$ (fully uncorrelated) where 
$\underline{a_{R}}$ is a vector of dimension $P$ with components $a_{R}^{\mu
\lbrack t]}$ for $\mu
\lbrack t]=1,2,...,P$, we obtain
\begin{eqnarray}
\sigma _{\Psi}^{2} &=&\sum_{R=1}^{2P}\left\langle
\left( n_{R}^{\underline{S}[t]}\right)
^{2}-n_{R}^{\underline{S}[t]}n_{\overline{R}}^{\underline{S}[t]}\right\rangle
_{t}+\sum_{R\neq R^{\prime }\neq \overline{R}}^{2P}\left\langle a_{R}^{\mu \lbrack
t]}a_{R^{\prime }}^{\mu \lbrack t]}\right\rangle _{t}\left\langle
n_{R}^{\underline{S}[t]}n_{R^{\prime }}^{
\underline{S}[t]}\right\rangle _{t} \nonumber \\ 
&=&\sum_{R=1}^{2P}\left\langle
\left( n_{R}^{\underline{S}[t]}\right)
^{2}-n_{R}^{\underline{S}[t]}n_{\overline{R}}^{\underline{S}[t]}\right\rangle
_{t}.  
\end{eqnarray} 
The sum over $2P$ terms can be written equivalently as a sum over
$P$ terms,
\begin{eqnarray}
\sigma _{\Psi}^{2} &=&\sum_{R=1}^{P}\left\langle
\left( n_{R}^{\underline{S}[t]}\right)
^{2}-n_{R}^{\underline{S}[t]}n_{\overline{R}}^{\underline{S}[t]}+\left(
n_{\overline{R}}^{\underline{S}[t]}\right)
^{2}-n_{\overline{R}}^{\underline{S}[t]}n_{R}^{\underline{S}[t]}\right\rangle _{t} 
\nonumber \\ 
&=&\sum_{R=1}^{P}\left\langle \left(
n_{R}^{\underline{S}[t]}-n_{\overline{R}}^{\underline{S}[t]}\right)
^{2}\right\rangle _{t} \equiv \left\langle \sum_{R=1}^{P}\left(
n_{R}^{\underline{S}[t]}-n_{\overline{R}}^{\underline{S}[t]}\right)
^{2}\right\rangle _{t}.  \nonumber
\end{eqnarray} The values of $n_{R}^{\underline{S}[t]}$ and
$n_{\overline{R}}^{\underline{S}[t]}$\ for each $R$ will depend on the precise form
of $\Psi $.  
The ensemble-average  over all possible
realizations of the strategy allocation matrix $\Psi$ is denoted by $\left\langle
...\right\rangle _{\Psi}$. Using the notation $\left\langle \sigma
_{\Psi}^{2}\right\rangle _{\Psi}=\sigma ^{2}$, yields
\begin{equation}
\sigma ^{2}=\left\langle \left\langle \sum_{R=1}^{P}\left(
n_{R}^{\underline{S}[t]}-n_{\overline{R}}^{\underline{S}[t]}\right)
^{2}\right\rangle _{t}\right\rangle _{\Psi } \ \ . \label{ModelC1}
\end{equation} 
It is straightforward to obtain analogous expressions for the variances in
$n_{+1}[t]$ and
$n_{-1}[t]$. 

Equation (\ref{ModelC1}) provides us with an
expression for the time-averaged fluctuations.  Some form of approximation must be
introduced in order to reduce Equation (\ref{ModelC1}) to explicit analytic
expressions. It turns out that Equation (\ref{ModelC1}) can be manipulated in a
variety of ways, depending on the level of approximation that one is prepared to
make. The precise form of any resulting analytic expression will depend on the
details of the approximations made. 
Adopting one such approach which is well-suited to the low $m$ regime, we start by
relabelling the strategies. Specifically, the sum in Equation
(\ref{ModelC1}) is re-written to be over a \emph{virtual-point ranking} $K$\ as
opposed to 
$R$. Consider the variation in points for a given strategy, as a function of time
for a given realization of $\Psi $. The ranking (i.e. label) of a given strategy
in terms of virtual-points score will typically change in time since the
individual strategies have a variation in virtual-points which also varies in
time. For the Minority Game, this variation is quite rapid in the low $m$ regime
of interest, since there are many more agents than available strategies -- hence
any strategy emerging as the instantaneously highest-scoring, will immediately get
played by many agents and therefore be likely to lose on the next time-step. More
general games involving competition within a multi-agent population, will typically
generate a similar ecology of strategy-scores with no all-time winner. 
This implies that the specific identity of the `$K$'th highest-scoring strategy'
changes frequently in time. It also implies that
$n_{R}^{\underline{S}[t]}$ varies considerably in time. Therefore in order to
proceed,  we shift the focus onto the time-evolution of the highest-scoring
strategy, second highest-scoring strategy etc. This should have a much smoother
time-evolution than the time-evolution for a given strategy. In
the case that the strategies all start off with zero points, the anticorrelated
strategies appear as the mirror-image, i.e.
$S_{K}[t]=-S_{\overline K}[t]$.  The label
$K$ is used to denote the rank in terms of strategy score, i.e.
$K=1$ is the highest scoring strategy position, $K=2$ is the second
highest-scoring strategy position etc. with
\begin{equation} S_{K=1}>S_{K=2}>S_{K=3}>S_{K=4}>...
\end{equation} 
assuming no strategy-ties. Given that
$S_{R}=-S_{\overline{R}}$\ (i.e. all strategy scores start off at zero), then we
know that
$S_{K}=-S_{\overline{K}}$. Equation (\ref{ModelC1}) can hence be rewritten exactly
as
\begin{equation}
\sigma^{2}=\left\langle \left\langle \sum_{K=1}^{P}\left(
n_{K}^{\underline{S}[t]}-n_{\overline{K}}^{\underline{S}[t]}\right)
^{2}\right\rangle _{t}\right\rangle _{\Psi }.  \label{ModelC1K}
\end{equation}  
Since in the systems of
interest the agents are typically playing their highest-scoring strategies, then
the relevant quantity in determining how many agents will instantanously play a
given strategy, is a knowledge of its relative ranking -- not the actual value of
its virtual points score. This suggests that the quantities
$n_{K}^{\underline{S}[t]}$\ and
$n_{\overline{K}}^{\underline{S}[t]}$\ will fluctuate relatively little in time,
and that we should now develop the problem in terms of time-averaged values. 
We can rewrite the number of agents playing the strategy in position $K$
at any timestep $t$, in terms of some constant value $n_K$ plus a fluctuating term
$n_{K}^{\underline{S}[t]}=n_{K}+\varepsilon _{K}[t]$. 
We assume that
one can choose a suitable constant $n_K$ such that the fluctuation $\varepsilon
_{K}[t]$ represents a small noise term. Hence,
\begin{eqnarray}
\sigma^{2} &=&\left\langle \sum_{K=1}^{P}\left\langle \left[ n_{K}+\varepsilon
_{K}[t]-n_{\overline{K}}-\varepsilon _{\overline{K}}[t]
\right] ^{2}\right\rangle _{t}\right\rangle _{\Psi } \label{average}\\
&\approx&\left\langle
\sum_{K=1}^{P}\left\langle \left[ n_{K}-n_{\overline{K}}\right] ^{2}\right\rangle
_{t}\right\rangle _{\Psi }=
\left\langle \sum_{K=1}^{P}\left[ n_{K}-n_{\overline{K}}\right] ^{2}\right\rangle
_{\Psi },  \nonumber
\end{eqnarray} assuming the noise terms have averaged out to be
small. The averaging over
$\Psi$ can now be taken inside the sum. Each term can then be rewritten
exactly using the joint probability distribution for $n_{K}$ and
$n_{\overline{K}}$, which we shall call
$P(n_{K},n_{\overline{K}})$. Hence
\begin{equation}
\sigma^{2} =\sum_{K=1}^{P}\left\langle \left[ n_{K}-n_{\overline{K}}\right]
^{2}\right\rangle _{\Psi}  \label{general} 
=\sum_{K=1}^{P}\sum_{n_{K}=0}^{N}\sum_{n_{\overline{K}}=0}^{N}
\left[ n_{K}-n_{\overline{K}}\right] ^{2}P(n_{K},n_{\overline{K}}) .  \nonumber
\end{equation}
We now look at
Equation (\ref{general}) in the limiting case where the averaging over the quenched
disorder matrix is dominated by matrices $\Psi$ which are nearly flat. This will
be a good approximation in the `crowded' limit of small $m$ in which there are
many more agents than available strategies, since the standard deviation of an
element in $\Psi$ (i.e. the standard deviation in bin-size) is then much smaller
than the mean bin-size. The probability distribution $P(n_{K},n_{\overline{K}})$
will then be sharply peaked around the $n_{K}$ and $n_{\overline{K}}$ values given
by the mean values for a flat quenched-disorder matrix $\Psi $. We label these
mean values as
${{\overline{n_{K}}}}$ and ${{\overline{n_{\overline{K}}}}}$.
Hence $P(n_{K},n_{\overline{K}})=
\delta_{n_{K},{{\overline{n_{K}}}}}\delta_{n_{\overline{K}},{\overline
{n_{\overline{K}}}}}$ and so
\begin{equation}
\sigma^{2}=\sum_{K=1}^{P}\left[ {\overline{n_{K}}}-{{\overline
{n_{\overline{K}}}}}\right] ^{2}.  \label{final}
\end{equation}  
There is a very simple interpretation of  Equation
(\protect\ref{final}). It represents the sum of the variances for each
Crowd-Anticrowd pair. For a given strategy $K$ there is an anticorrelated strategy
$\overline{K}$. The ${\overline{n_{K}}}$ agents using strategy $K$\ are doing the
\emph{opposite} of the ${{\overline {n_{\overline{K}}}}}$\ agents using strategy
$\overline{K}$\
\emph{irrespective} of the history bit-string. Hence the effective group-size for
each Crowd-Anticrowd pair is
$n_K^{eff}={\overline{n_{K}}}-{{\overline {n_{\overline{K}}}}}\ $ : this
represents the net step-size
$d$ of the Crowd-Anticrowd pair in a random-walk contribution to the total
variance. Hence, the net contribution by this Crowd-Anticrowd pair to the variance
is given by
\begin{equation} [\sigma^{2}]_{K\overline{K}} = 4pqd^{2} =
4.\frac{1}{2}.\frac{1}{2}[n_K^{eff}]^{2}= 
\left[{\overline{n_{K}}}-{{\overline {n_{\overline{K}}}}}\right]^{2}
\end{equation} where $p=q=1/2$ for a random walk. Since all the strong
correlations have been included (i.e. anti-correlations) it can therefore be
assumed that the separate Crowd-Anticrowd pairs execute random walks which are
\emph{uncorrelated} with respect to each other. [Recall the properties of the RSS
- all the remaining strategies are uncorrelated.] Hence the total variance is
given by the sum of the individual variances,
\begin{equation}
\sigma^{2}=\sum_{K=1}^{P}[\sigma^{2}]_{K\overline{K}}=\sum_{K=1}^{P}\left[
{\overline{n_{K}}}-{{\overline {n_{\overline{K}}}}}\right] ^{2},
\end{equation} which corresponds exactly to Equation (\ref{final}). If
strategy-ties occur frequently, then one has to be more careful about evaluating
${\overline{n_{K}}}$ since its value may be affected by the tie-breaking rule. We
will show elsewhere that this becomes quite important in the case of very small
$m$ in the presence of network connections \cite{sean} - this is because very small
$m$ naturally leads to crowding in strategy space and hence mean-reverting virtual
scores for strategies. This mean-reversion is amplified further by the presence of
network connections which increases the crowding, thereby increasing the chance of
strategy ties.

\begin{figure}
\includegraphics[width=.4\textwidth]{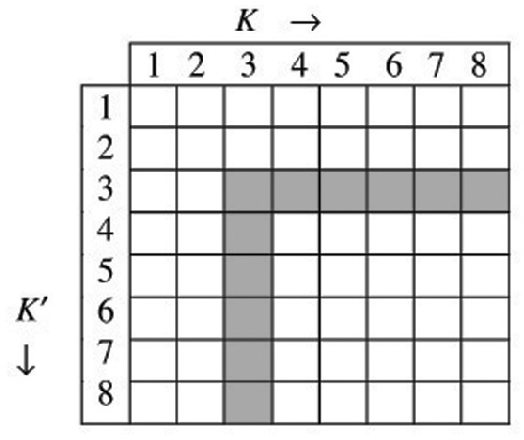}
\caption{Schematic representation of the strategy allocation matrix
$\Psi$ with $m=2$ and $S=2$, in the RSS. The strategies are ranked according to
strategy score, and are labelled by the rank $K$. In the limit that $\Psi$ is
essentially flat, then the number of agents playing the $K$'th highest-scoring
strategy, is just proportional to the number of shaded bins at that $K$.}
\label{figure3}
\end{figure}

\section{Implementation of Crowd-Anticrowd Theory}  Here we
evalute the Crowd-Anticrowd expressions, in the important
limiting case of small
$m$. Since there are  many more agents than available strategies,
crowding effects will be important. 
Each element of
$\Psi$ has a mean of
$N/(2P)^{S}$ agents per `bin'. In the case of small $m$ and hence densely-filled
$\Psi$, the fluctuations in the number of agents per bin will be small compared to
this mean value -- hence the matrix $\Psi$ looks uniform or `flat' in terms of
the occupation numbers in each bin.  Figure 3 provides a schematic representation
of
$\Psi$ with $m=2$, $S=2$, in the RSS. If the matrix $\Psi$ is indeed flat, then
any  re-ordering due to changes in the strategy ranking has no effect on the form
of the matrix. Therefore the number of agents playing the $K$'th highest-scoring
strategy, will always be proportional to the number of shaded bins at that $K$
(see Fig. 3 for $K=3$). 
For general $m$ and $S$, one finds
\begin{eqnarray}
{\overline{n_{K}}} &=&\frac{N}{(2P)^{S}}[S(2P-K)^{S-1}+\frac{S(S-1)}{2}
(2P-K)^{S-2}+...+1]  \label{YofR} \\
&=&\frac{N}{(2P)^{S}}\sum_{r=0}^{S-1}\frac{S!}{(S-r)!r!}[2P-K]^{r}  \nonumber \\
&=&\frac{N}{(2P)^{S}}([2P-K+1]^{S}-[2P-K]^{S})  \nonumber \\ &=&N.\left( \left[
1-\frac{(K-1)}{2P}\right] ^{S}-\left[ 1-\frac{K}{2P}
\right] ^{S}\right) ,  \nonumber
\end{eqnarray}
with $P\equiv 2^{m}$. In the case where each agent holds two
strategies, $S=2,$ ${\overline{n_{K}}}$ can be simplified to
\begin{equation} {\overline{n_{K}}} = N.\left( \left[ 1-\frac{(K-1)}{2P}\right]
^{2}-\left[ 1-
\frac{K}{2P}\right] ^{2}\right)   = \frac{(2^{m+2}-2K+1)}{2^{2(m+1)}}N\ . 
\label{k} 
\end{equation} 
Hence
\begin{eqnarray}
\sigma^{2} &=&\sum_{K=1}^{P}\left[ \frac{(2^{m+2}-2K+1)}{
2^{2(m+1)}}N-\frac{(2K-1)}{2^{2(m+1)}}N\right] ^{2} \\
&=&\frac{N^{2}}{2^{2(2m+1)}}\sum_{K=1}^{P}[2^{m+1}-2K+1]^{2} 
=\frac{N^{2}}{3.2^{m}}(1-2^{-2(m+1)})\ \ . \nonumber
\end{eqnarray} 
This derivation has
assumed that there are no strategy ties -- more precisely, we have assumed that
the game rules governing strategy ties do not upset the identical forms of the
rankings in terms of highest virtual points and popularity. Hence we have
overestimated the size of the Crowds using high-ranking strategies, and
underestimated the size of the Anticrowds using low-ranking strategies. Therefore
the analytic form for $\sigma$ will overestimate the numerical value, as is indeed 
seen in Figure 4. Notwithstanding this overestimation, there is remarkably good
agreement between the numerical results and our analytic theory. In a similar way
to the above calculation, the Crowd-Anticrowd theory can be extended to deal with
the important complementary regimes of (i) non-flat quenched disorder matrix
$\Psi$, at small
$m$, and (ii) non-flat quenched disorder matrix $\Psi$, at large $m$. As shown in
Figure 4, the agreement for these regimes is also excellent \cite{book,lanl}.

\begin{figure}
\includegraphics[width=.7\textwidth]{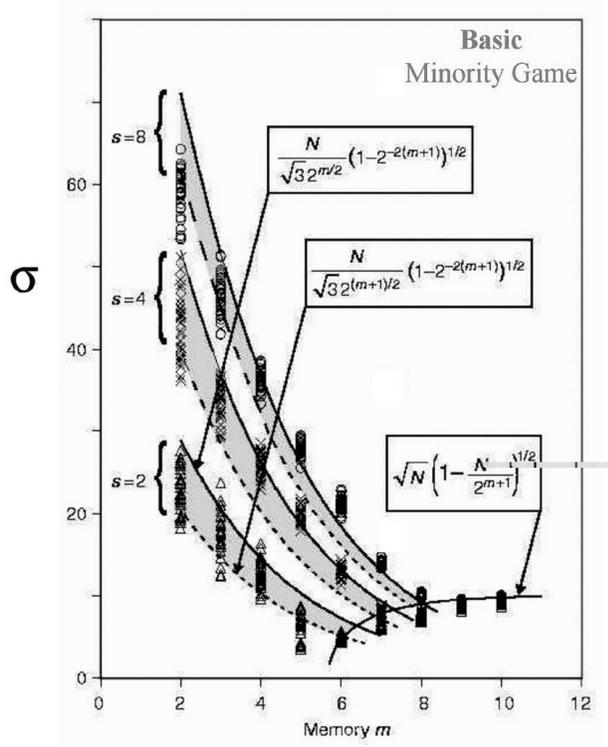}
\caption{Crowd-Anticrowd theory vs. numerical simulation results for $\sigma$ in
the Minority Game  as a function of memory size $m$, for $N=101$ agents, at $S=2$,
$4$ and $8$. At each $S$ value, analytic forms of $\sigma$ (i.e. standard deviation
in excess demand
$D[t]$) are shown. The
numerical values were obtained from different simulation runs (triangles, crosses
and circles).  Figure adapted from Ref. \cite{book}.}
\label{figure4}
\end{figure}

The Crowd-Anticrowd theory has also been applied successfully to various 
generalizations of the Minority Game. For example, excellent agreement between
the resulting analytic expressions and numerical simulations has been demonstrated
for (i) Alloy Minority Game
\cite{alloy}, (ii) Thermal Minority Game (TMG) \cite{SherThermLett,generalTherm},
(iii) Thermal Alloy Minority Game
\cite{CAA}, and (iv) B-A-R systems with an underlying network structure
\cite{lanl}. 

\section{Conclusion and Discussions} We have given an overview of the
Crowd-Anticrowd theory for competitive multi-agent systems,
in particular those based on an underlying binary structure. Explicit analytic
expressions can be evaluated at various levels of
approximation, yielding very good agreement with numerical simulations. We note
that the crucial element of this Crowd-Anticrowd theory -- i.e. properly
accounting for the dominant inter-agent correlations -- is not limited to one
specific game. Given its success in describing a number of generalized B-A-R
systems, we believe that the Crowd-Anticrowd framework could provide a powerful
approach to describing a wide class of Complex Systems which mimic competitive
multi-agent games. This would be a welcome development, given the lack of general
theoretical concepts in the field of Complex Systems as a whole. It is also
pleasing from the point of view of physics methodology, since the basic underlying
philosophy of accounting correctly for `inter-particle' correlations is already
known to be successful in more conventional areas of many-body physics. This
success in turn raises the intriguing possibility that conventional many-body
physics might be open to re-interpretation in terms of an appropriate
multi-particle `game': we leave this for future work.

Of course, some properties of Complex Systems cannot be described using time- and
configuration-averaged expressions as discussed here. In particular, an observation
of a real-world Complex System which is thought to resemble a multi-agent game, may
correspond to a
\emph{single} run which evolves from a specific initial configuration of agents'
strategies. This implies a particular $\Psi$, and hence the time-averagings within
the Crowd-Anticrowd theory must be carried out for that particular choice of
$\Psi$. However this problem can still be cast in terms of the Crowd-Anticrowd
approach, since the averagings are then just carried out over some sub-set of
paths in history space, which is conditional on the path along which the Complex
System is already heading. 

We have been discussing a Complex System based on multi-agent dynamics, in which
both deterministic and stochastic processes co-exist, and are indeed intertwined.
Depending on the particular rules of the game, the stochastic element may be
associated with any of five areas: (i) disorder associated with the strategy
allocation and hence with the heterogeneity in the population, (ii) disorder in
an underlying network. Both (i) and (ii) might typically be fixed from the outset
(i.e., quenched disorder) hence it is interesting to see the interplay of (i) and
(ii) in terms of the overall performance of the system \cite{sean}. The extent to
which these two `hard-wired' disorders might then compensate each other, as for
example in the Parrondo effect or stochastic resonance, is an interesting
question. Such a compensation effect might be engineered, for example, by altering
the rules-of-the-game concerning inter-agent communication on the existing network.
Three further possible sources of stochasticity are (iii) tie-breaks in the scores
of strategies, (iv) a stochastic rule in order for each agent to pick which
strategy to use from the available $S$ strategies, as in the Thermal Minority
Game, (v) stochasticity in the global resource level $L[t]$ (e.g. bar seating
capacity) due to changing external conditions. To a greater or lesser extent, these
five stochastic elements will tend to break up any deterministic cycles arising in
the game. We refer to Ref.
\cite{paul} for a discussion of the dynamics of the Minority Game viewed from the
perspective of a stochastically-perturbed deterministic system.

%

\end{document}